\documentclass[10pt,preprint]{aastex}

\def\lens{J0134--0931}
\def\ltsima{$\; \buildrel < \over \sim \;$}
\def\lsim{\lower.5ex\hbox{\ltsima}}
\def\gtsima{$\; \buildrel > \over \sim \;$}
\def\gsim{\lower.5ex\hbox{\gtsima}}
\newcommand\cp{${\rm C}'$}
\newcommand\ep{${\rm E}'$}

\slugcomment{}

\shorttitle{Quintuple Quasar:\ Observations}
\shortauthors{Winn et al.}

\begin{document}

\title{The Quintuple Quasar: Radio and Optical Observations}

\author{
Joshua N.\ Winn\altaffilmark{1,2},
Christopher S.\ Kochanek\altaffilmark{2},
Charles R.\ Keeton\altaffilmark{3,4},
James E.J.\ Lovell\altaffilmark{5}
}

\email{jwinn@cfa.harvard.edu;
ckochanek@cfa.harvard.edu;
ckeeton@oddjob.uchicago.edu;
Jim.Lovell@atnf.csiro.au}

\altaffiltext{1}{NSF Astronomy \& Astrophysics Postdoctoral Fellow}

\altaffiltext{2}{Harvard-Smithsonian Center for Astrophysics, 60
Garden St., Cambridge, MA 02138}

\altaffiltext{3}{Department of Astronomy and Astrophysics, University
of Chicago, 5640 South Ellis Avenue, Chicago, IL 60637}

\altaffiltext{4}{Hubble fellow}

\altaffiltext{5}{CSIRO Australia Telescope National Facility, P.O.\
Box 76, Epping, NSW 1710, Australia}

\begin{abstract}
We present results from high-resolution radio and optical observations
of PMN~J0134--0931, a gravitational lens with a unique radio
morphology and an extremely red optical counterpart.  Our data support
the theory of Keeton \& Winn (2003): five of the six observed radio
components are multiple images of a single quasar, produced by a pair
of lens galaxies.  Multi-frequency VLBA maps show that the sixth and
faintest component has a different radio spectrum than the others,
confirming that it represents a second component of the background
source rather than a sixth image.  The lens models predict that there
should be additional faint images of this second source component, and
we find evidence for one of the predicted images.  The
previously-observed large angular sizes of two of the five bright
components are not intrinsic (which would have excluded the
possibility that they are lensed images), but are instead due to
scatter broadening.  Both the extended radio emission observed at low
frequencies, and the intrinsic image shapes observed at high
frequencies, can be explained by the lens models.  The pair of lens
galaxies is marginally detected in HST images.  The differential
extinction of the quasar images suggests that the extreme red color of
the quasar is at least partly due to dust in the lens galaxies.
\end{abstract}


\section{Introduction}
\label{sec:intro}

The quasar PMN~J0134--0931 calls attention to itself in many ways.  It
is very bright at radio frequencies (1~Jy at 2~GHz); its optical
counterpart is extraordinarily red ($B-K\geq 11$); and it has a unique
radio morphology consisting of six compact components within a circle
of diameter $0\farcs7$.  It is therefore unsurprising, in retrospect,
that it was discovered to be an interesting object by two independent
groups.

Winn et al.~(2002a; hereafter, W02) found it during a survey for
radio-loud gravitational lenses.  They showed that at least five of
the six radio components have identical continuum radio spectra, and
discovered a curved arc of radio emission joining two of the
components.  These properties are hallmarks of strong gravitational
lensing.

Gregg et al.~(2002; hereafter, G02) found it in a cross-compilation of
radio, optical, and near-infrared catalogs that was designed to
identify the reddest quasars on the sky.  They measured its redshift
($z_s=2.2$) and found at least three components in a near-infrared
image.  The multiplicity of the near-infrared counterpart, and the
extremely large inferred luminosity ($M_R=-29.6$, even under the
conservative assumption of no extinction) were evidence of
gravitational lensing.

However, neither group could explain the unique morphology with a
plausible lens model, because strong lensing by a single galaxy
usually results in only two or four images.  Another difficulty was
that two of the components were found to have a lower surface
brightness than the other components in high-resolution radio images,
in apparent contradiction of the conservation of surface brightness by
gravitational lensing.

Also unknown was the reason for the extremely red color of \lens.  The
two most obvious possibilities---reddening due to dust in the host
galaxy, or due to dust in the lens galaxy---could not be
distinguished.  \citet{hall02} detected strong \ion{Ca}{2}~H and K
absorption lines at $z_l=0.76$ (presumably the lens redshift), but
they were not able to determine the reason for the redness of the
quasar spectrum.

In a companion paper, Keeton \& Winn (2003; hereafter, KW03) present
the first quantitative models that explain all the previously known
properties of \lens.  In those models, five of the radio components
are multiple images of a single quasar, and the sixth component
represents a different source, which is presumably a second component
of the same background radio source.  The positions of the two lens
galaxies are constrained remarkably well by the observed image
configuration.  Furthermore, the two low-surface-brightness components
are located close to the expected position of one of the lens
galaxies, suggesting that scatter broadening by ionized material in
that lens galaxy could account for their large angular sizes.

In addition to explaining the previous data, the KW03 models make a
number of predictions about more subtle properties of \lens\ that
would be detectable in radio and optical observations with high
sensitivity and high angular resolution.  In this paper we present
multi-frequency observations of \lens\ with the Very Long Baseline
Array (VLBA\footnote{The Very Long Baseline Array (VLBA) is operated
by the National Radio Astronomy Observatory, a facility of the
National Science Foundation operated under cooperative agreement by
Associated Universities for Research in Astromomy, Inc.}) and the
Hubble Space Telescope (HST\footnote{Data from the {\sc nasa/esa}
Hubble Space Telescope (HST) were obtained from the Space Telescope
Science Institute, which is operated by {\sc aura}, Inc., under {\sc
nasa} contract NAS~5-26555.}) that confirm some of these predictions,
and refute none of them.

Before describing the new observations, in \S~\ref{sec:overview} we
provide an overview of the previously-known radio morphology of \lens,
and the nomenclature used to describe it. We also review the main
features of the KW03 models.  The radio data (\S~\ref{sec:radio})
confirm that the sixth and faintest radio component has a
significantly different spectral index than the other five components
(\S~\ref{sec:cmptf}).  The two components that were observed to have
larger angular sizes are indeed being scatter broadened
(\S~\ref{sec:scatter}).  There is tantalizing (but tentative) evidence
for one of the additional images of the second background source that
are predicted by the KW03 models (\S~\ref{sec:counterimage}).  The
extended radio emission (\S~\ref{sec:arc}) and component shapes
(\S~\ref{sec:shapes}) can be explained by the models.

After subtracting point sources from the optical images
(\S~\ref{sec:optical}), we identify two faint but significant peaks of
residual flux as direct detections of the lens galaxies
(\S~\ref{sec:lensgals}).  The optical counterparts of the radio
components have different colors, and their optical flux ratios differ
from their radio flux ratios, implying that the extreme red color of
\lens\ is at least partly due to reddening by dust in the lens galaxy
(\S~\ref{sec:colors}).  Finally, in \S~\ref{sec:summary}, we summarize
the evidence that \lens\ is the first known instance of a
quintuply-imaged quasar.

\section{Overview}
\label{sec:overview}

\begin{figure}[h]
\figurenum{1}
\epsscale{0.7}
\plotone{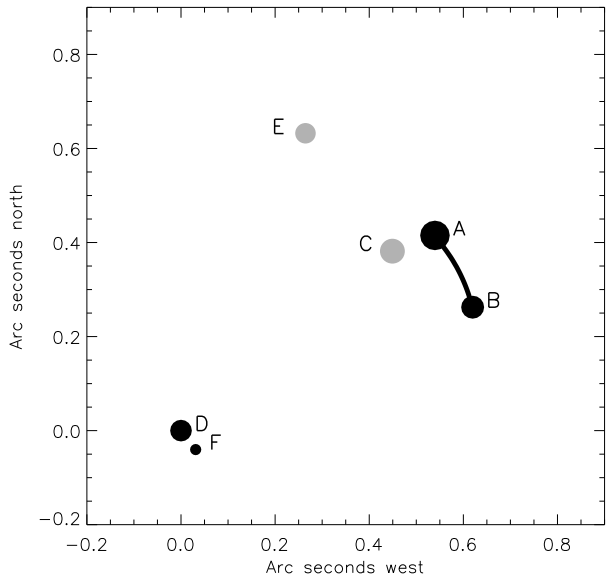}
\caption{Illustration of the morphology of \lens.
\label{fig:overview}}
\end{figure}

Figure~\ref{fig:overview} is a diagram of the observed configuration
of the radio components.  The dots represent the six compact radio
components A--F, which are named in decreasing order of radio flux
density.  The size of each dot is proportional to the logarithm of its
flux density ratio relative to component D.  The angular sizes of C
and E are much larger than the other components at 1.7~GHz and
5.0~GHz, which is why we have used gray dots to represent those
components instead of black dots.  There is a curved arc extending
between A and B at 1.7~GHz, represented schematically by a solid line.
Components A--E all have the same spectral index, as measured from
5~GHz to 43~GHz ($\alpha = -0.69\pm 0.04$, where $S_\nu \propto
\nu^{\alpha}$), but the spectral index of F was unknown.

\begin{figure}[h]
\figurenum{2}
\epsscale{1.0}
\plottwo{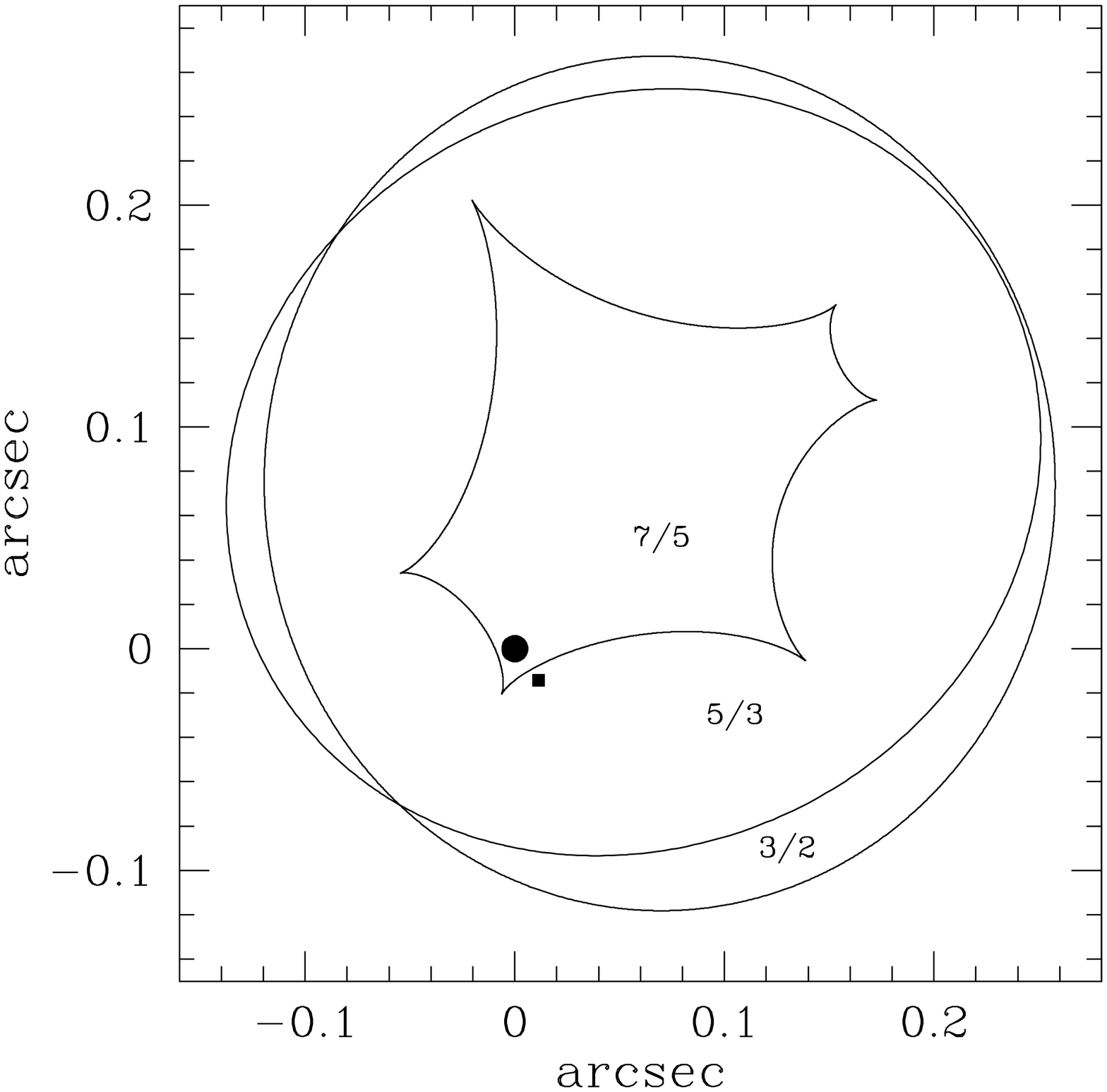}{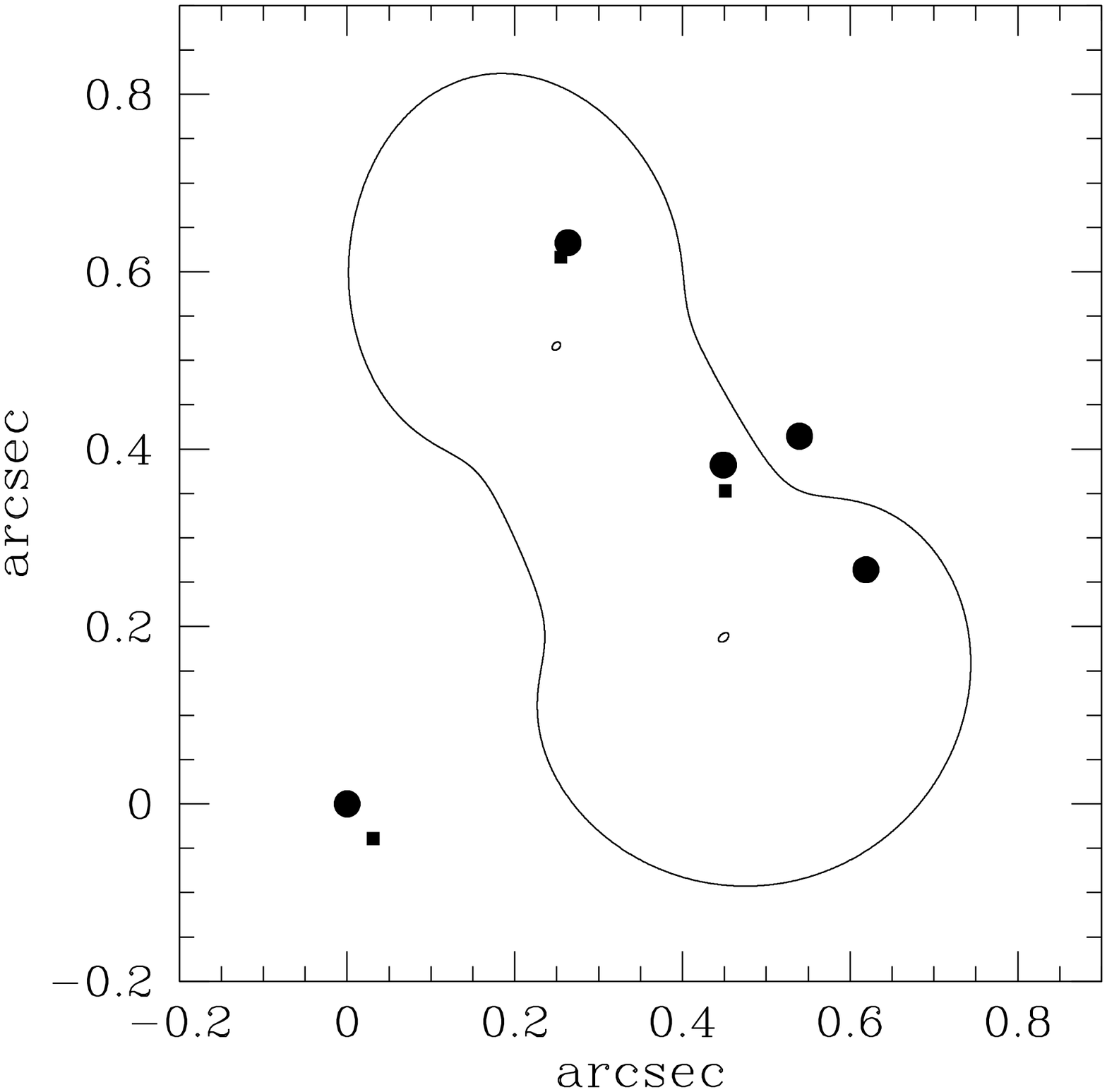}
\caption{Representative lens model of \lens, from KW03.  {\em Left:}
source plane, with caustics (solid lines), source component S$_1$
(black dot), and source component S$_2$ (black square).  The image
multiplicity of each region is reported as $N_{\rm tot}/N_{\rm bri}$,
where $N_{\rm tot}$ is the total number of images, and $N_{\rm bri}$
is the number of bright (non-core) images.  {\em Right:} image plane,
with critical curves (solid lines), images of S$_1$ (black dots), and
images of S$_2$ (black squares).
\label{fig:lensmodel}}
\end{figure}

According to the KW03 models, the background radio source has two
components (S$_1$ and S$_2$) and there are two lens galaxies (Gal-N
and Gal-S).  The five observed radio components A--E are five images
of S$_1$, and F is the brightest of three images of S$_2$.  The
positions and masses of the lens galaxies are fairly well constrained
by the data, but their ellipticities and position angles are not.  For
reference, the critical curves and caustics of a representative lens
model are shown in Fig.~\ref{fig:lensmodel}.  The reader is referred
to KW03 for a full theoretical analysis.

\section{Radio evidence}
\label{sec:radio}

\subsection{Observations and calibration}
\label{sec:calibration}

This system was observed with the VLBA at 1.7~GHz and 5.0~GHz by W02.
Below, we present the results from new VLBA observations at three
different frequencies.  We also present images based on both the old
and the new data, to provide a uniform multi-frequency study.  A list
of the basic parameters of all the VLBA observations is given in
Table~\ref{tbl:journal}.

The new observations were conducted in standard VLBA continuum
frequency bands at 2.3~GHz, 8.4~GHz, and 15~GHz.  The array consisted
of the usual ten antennas, except that the Pie Town antenna did not
participate in the 15~GHz observation.\footnote{Instead, one antenna
from the Very Large Array (VLA) participated. However, the data from
this antenna were much noisier than the VLBA data, and we did not
include data from the VLA antenna in our final analysis.}  For the
2.3~GHz observation, we did not employ phase referencing because the
source was sufficiently bright for self-calibration.  At the higher
frequencies, we switched between \lens\ and the nearby, bright,
compact source J0141--0928 in order to calibrate the relative antenna
gains.  The cycle times were 4 minutes and 2 minutes at 8.4~GHz and
15~GHz, respectively.  Short observations of the bright radio sources
J2253+1608 and J0555+3948 were included to serve as fringe-finders and
to correct large delay errors.

The observing bandwidth of 32~MHz per polarization was divided into 4
sub-bands of width 8~MHz.  Both senses of polarization were recorded
with 1-bit sampling.  The data were correlated in Socorro, New Mexico,
producing 16 channels of width 0.5~MHz from each sub-band, with an
integration time of one second.

Calibration was performed with {\sc aips}\footnote{The Astronomical
Image Processing System ({\sc aips}) is developed and distributed by
the National Radio Astronomy Observatory (NRAO).} using standard
procedures summarized as follows.  Obviously corrupted data, and data
taken at elevations less than $15\arcdeg$, were discarded.  Visibility
amplitudes were calibrated using on-line measurements of antenna
gains, system temperatures, and voltage offsets in the samplers.
Large delay errors were removed by fringe-fitting the data from a
1.5-minute observation of J2253+1608 and applying the delay
corrections to all the data.  At 2.3~GHz, we corrected residual rates,
delays, and phases by fringe-fitting the data from \lens\ directly; at
the higher frequencies we fringe-fitted data from J0141--0928 and
interpolated the solutions.  After calibration, the data were averaged
in time and frequency in order to reduce the data volume, but we
ensured that the final sampling was still fine enough to produce
$<$1\% smearing over the desired field of view.

\begin{deluxetable}{lccrccrc}
\tabletypesize{\tiny}
\tablecaption{Observations of \lens\ with the VLBA.\label{tbl:journal}}
\tablewidth{0pt}
\tablehead{
\colhead{Start Date} &
\colhead{Frequency} &
\colhead{Duration} &
\colhead{Field size} &
\colhead{Pixel scale} &
\multicolumn{2}{c}{Beam parameters} &
\colhead{R.M.S.\ noise} \\
\colhead{(UT)} &
\colhead{(GHz)} &
\colhead{(hours)} &
\colhead{($N_{\rm pix}\times N_{\rm pix}$)} &
\colhead{(mas~pixel$^{-1}$)} &
\colhead{$b_{\rm maj} \times b_{\rm min}$ (mas)} &
\colhead{P.A.\ (degrees)} &
\colhead{(mJy~beam$^{-1}$)}
}

\startdata
2000~Apr~28 &  4.99 & 1 &      $2048\times 2048$ & 0.50 & $ 6.5\times3.5$ & $ 12\fdg7$ & 0.30 \\
2000~Oct~31 &  1.67 & 4 &      $1024\times 1024$ & 1.00 & $17.1\times4.9$ & $ 19\fdg1$ & 0.20 \\
2001~Oct~17 &  2.27 & 7 &      $1024\times 1024$ & 1.00 & $10.1\times3.8$ & $ -3\fdg0$ & 0.23 \\
2001~Dec~08 &  8.42 & 7 &      $4096\times 4096$ & 0.25 & $ 4.3\times2.3$ & $  9\fdg8$ & 0.13 \\
2002~Feb~02 & 15.36 & 7 & ABC:~$2048\times 2048$ & 0.15 & $ 1.5\times0.5$ & $ -6\fdg3$ & 0.36 \\
            &       &   &  DF:~$1024\times 0.15$ & 0.15 & $ 1.5\times0.5$ & $ -6\fdg3$ & 0.32 \\
            &       &   &   E:~$1024\times 0.15$ & 0.15 & $ 1.5\times0.5$ & $ -6\fdg3$ & 0.37 \\
\enddata
\end{deluxetable}

\subsection{Images and general results}
\label{sec:images}

Imaging was performed with {\sc aips}.  For each frequency, we first
applied a Gaussian tapering function to the visibility data, to
emphasize the low-resolution (short-baseline) information, and used
the {\sc clean} algorithm to create a preliminary model of the
detected radio structures.  This model was used to self-calibrate the
antenna phases.  A new map was created with no tapering function, and
the process of {\sc clean}ing and self-calibration was repeated until
no further improvement was noted.  We then used the latest {\sc clean}
model to self-calibrate the relative amplitudes of the antennas with a
solution interval of 30~minutes.  The amplitude corrections were
typically smaller than 10\%.  We used a moderately uniform weighting
({\sc robust}~$=0$, in {\sc aips}) for the final images.

We also applied the self-calibration and imaging procedures described
above to the previously published 1.7~GHz and 5.0~GHz data, for the
sake of uniformity. This resulted in an improvement in dynamic range
over the previously published images, due to the increased care in
removing corrupted data, the use of amplitude self-calibration, and
the different choice of visibility weighting.

In all cases except 15~GHz, we created one wide-field image, centered
between components B and D in right ascension, and between D and E in
declination.  At 15~GHz we used the multiple-field {\sc clean}
algorithm to deconvolve three fields simultaneously: one field
centered on components A, B, and C; one field centered on component E;
and one field centered on components D and F.

\begin{figure}
\figurenum{3}
\epsscale{1.0}
\plotone{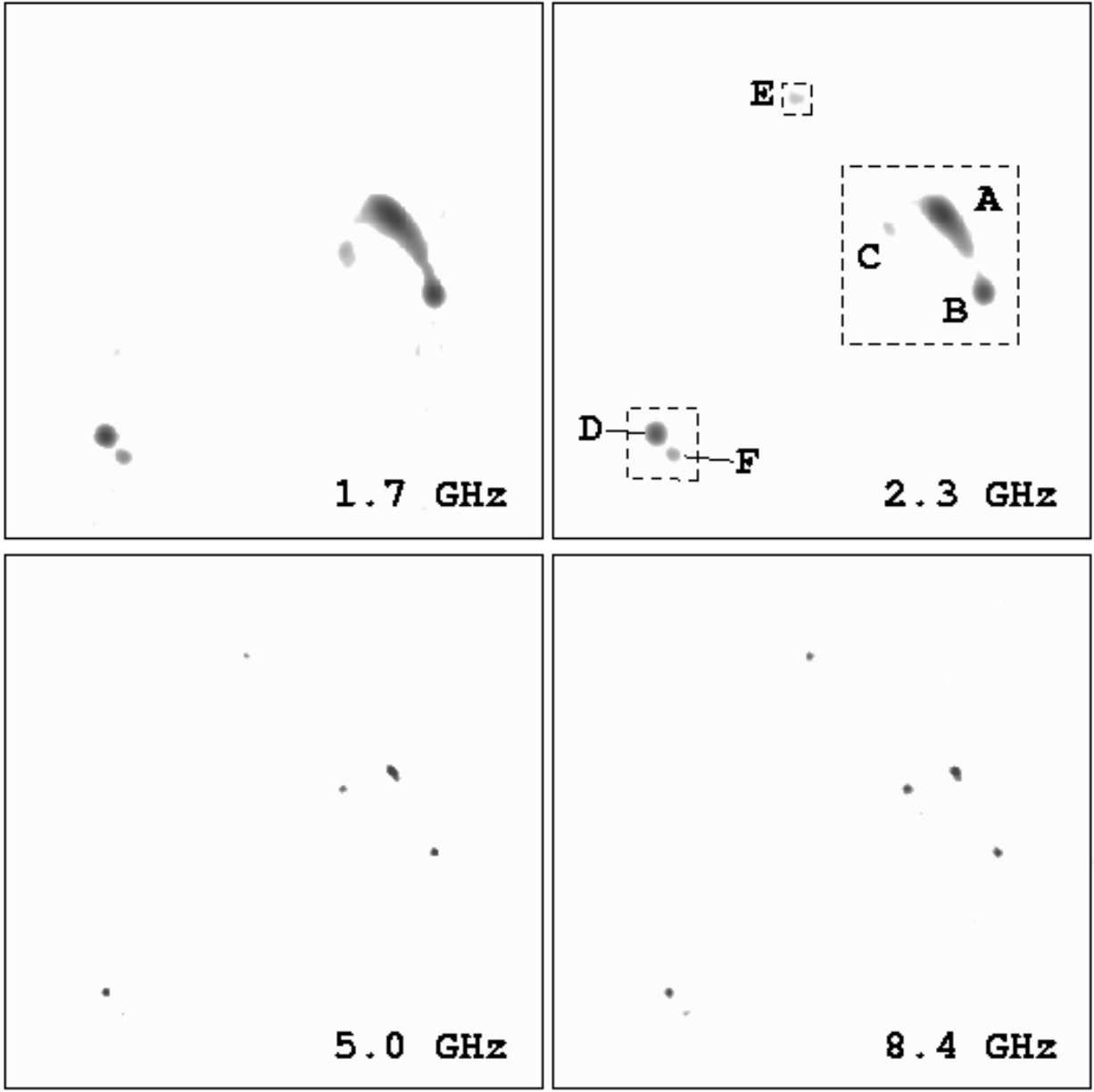}
\caption{Grayscale VLBA images ($1\arcsec\times 1\arcsec$) at four
frequencies.  The first gray level has been set to $8\sigma$, where
$\sigma$ is the R.M.S.\ noise level given in Table~\ref{tbl:journal}.
North is up and east is left.  In the upper right panel, A--F are
labeled, and dashed lines indicate the regions shown at full
resolution for 1.7--8.4~GHz in Figs.~\ref{fig:abc}--\ref{fig:e}.
\label{fig:grayscale-maps}}
\end{figure}

Figure~\ref{fig:grayscale-maps} shows wide-field grayscale images for
all frequencies except 15~GHz. These images provide an overview of the
detected radio structures.  To ease the visual interpretation of these
images, they were created after applying a Gaussian tapering function
to the visibility data, and were restored with a circular Gaussian
beam. For the 1.7~GHz and 2.3~GHz images, the tapering function
declined to 30\% at a visibility radius of $10^6\lambda$, and the
restoring beam had a diameter of 20~milli-arc~seconds (mas).  For the
5.0~GHz and 8.4~GHz images, the corresponding values were
$3\times10^6\lambda$ and 7~mas.

\begin{figure}
\figurenum{4}
\epsscale{1.0}
\plotone{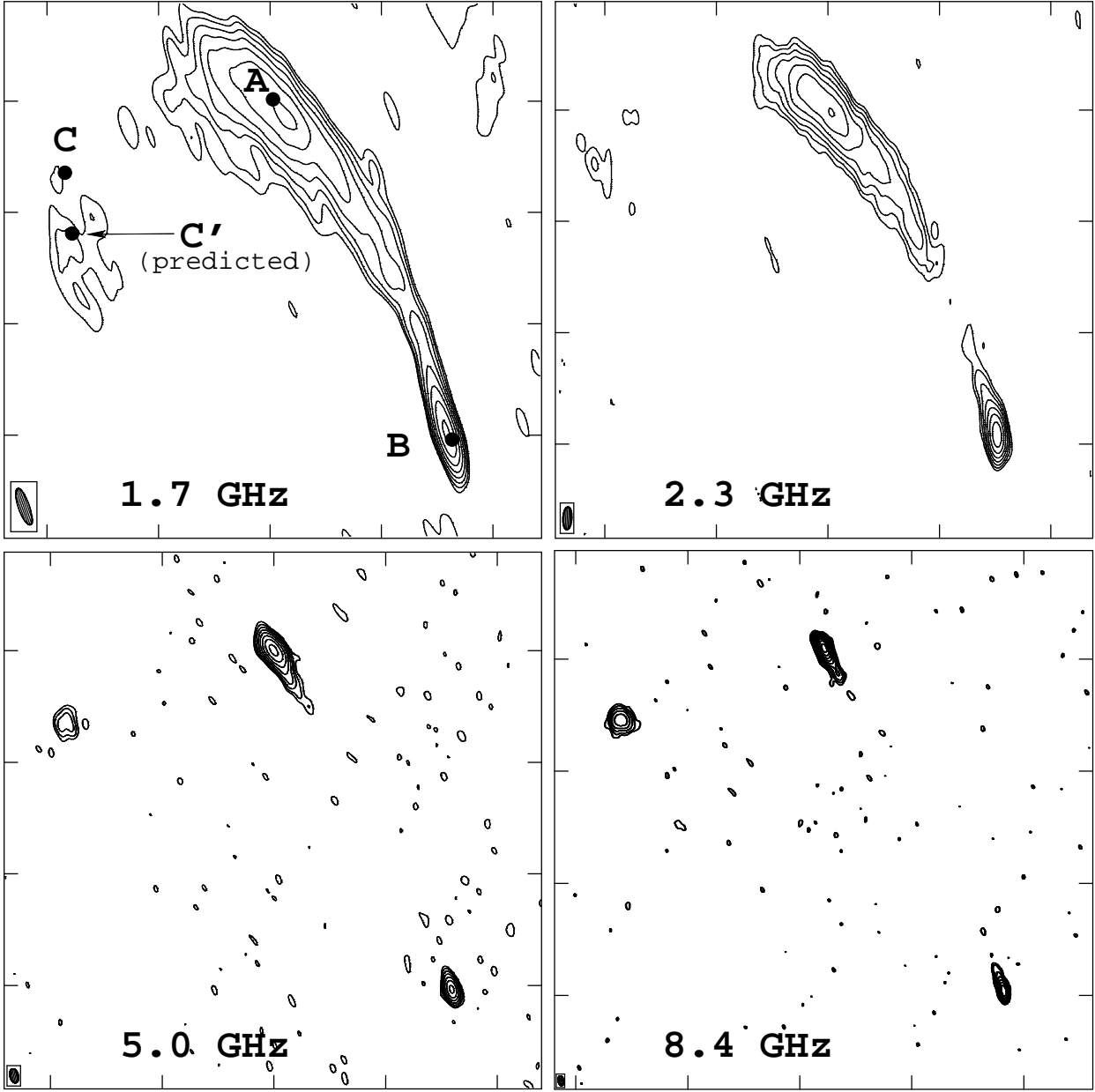}
\caption{Radio contour maps of the $0\farcs24\times 0\farcs24$ field
surrounding components A, B, and C.  Contours begin at $3\sigma$ and
increase by factors of 2, where $\sigma$ is the R.M.S.\ noise level.
The restoring beam is illustrated in the lower left corner of each
panel.  In the upper left panel (1.7~GHz), black dots mark the
positions of A, B, and C (as measured at higher frequencies) and the
predicted position of \cp.
\label{fig:abc}}
\end{figure}

\begin{figure}
\figurenum{5}
\plotone{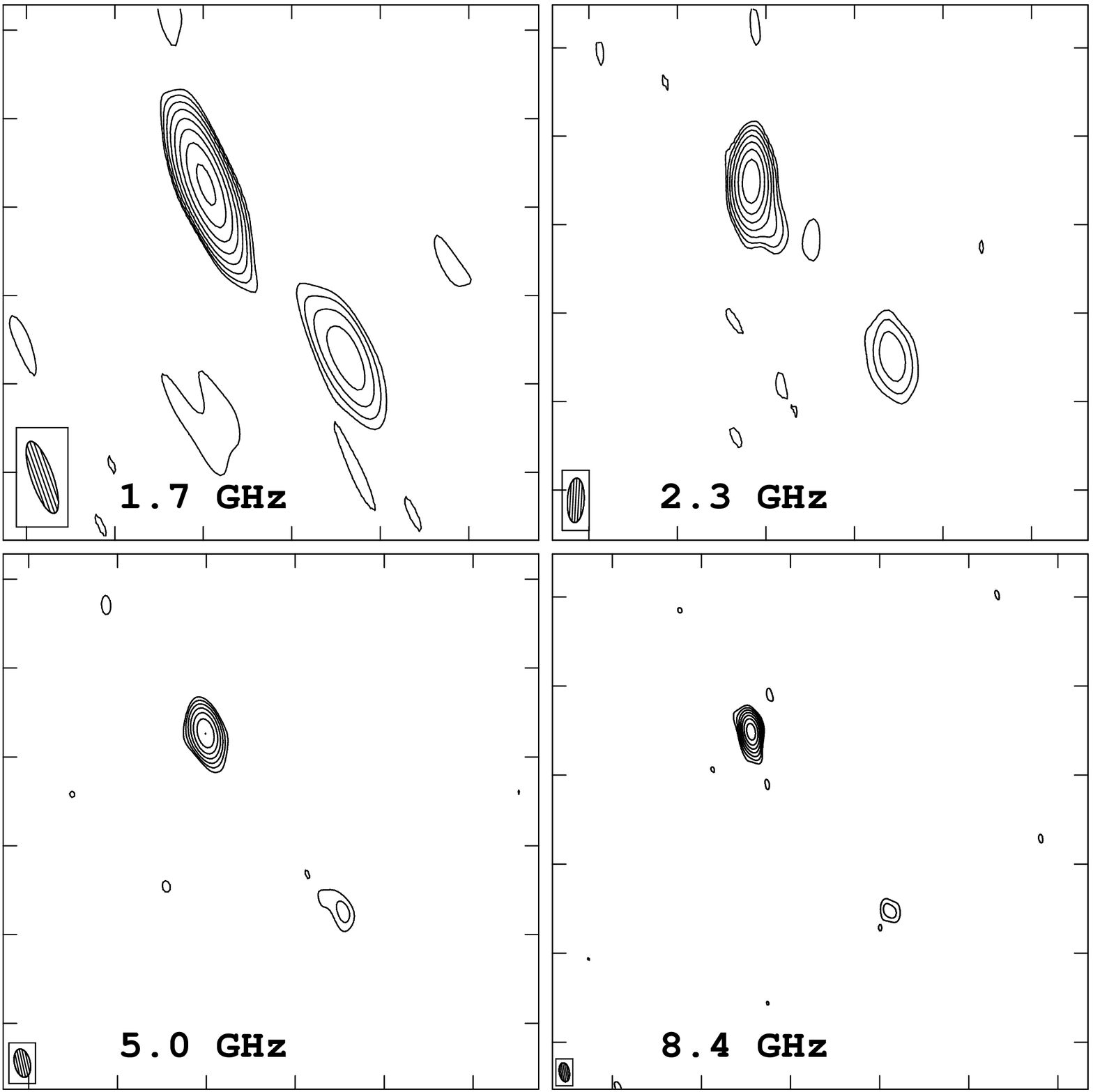}
\caption{Radio contour maps of the $0\farcs12\times 0\farcs12$ field
surrounding components D and F.  Contours begin at $3\sigma$ and
increase by factors of 2, where $\sigma$ is the R.M.S.\ noise level.
The restoring beam is illustrated in
the lower left corner of each panel.
\label{fig:df}}
\end{figure}

\begin{figure}
\figurenum{6}
\plotone{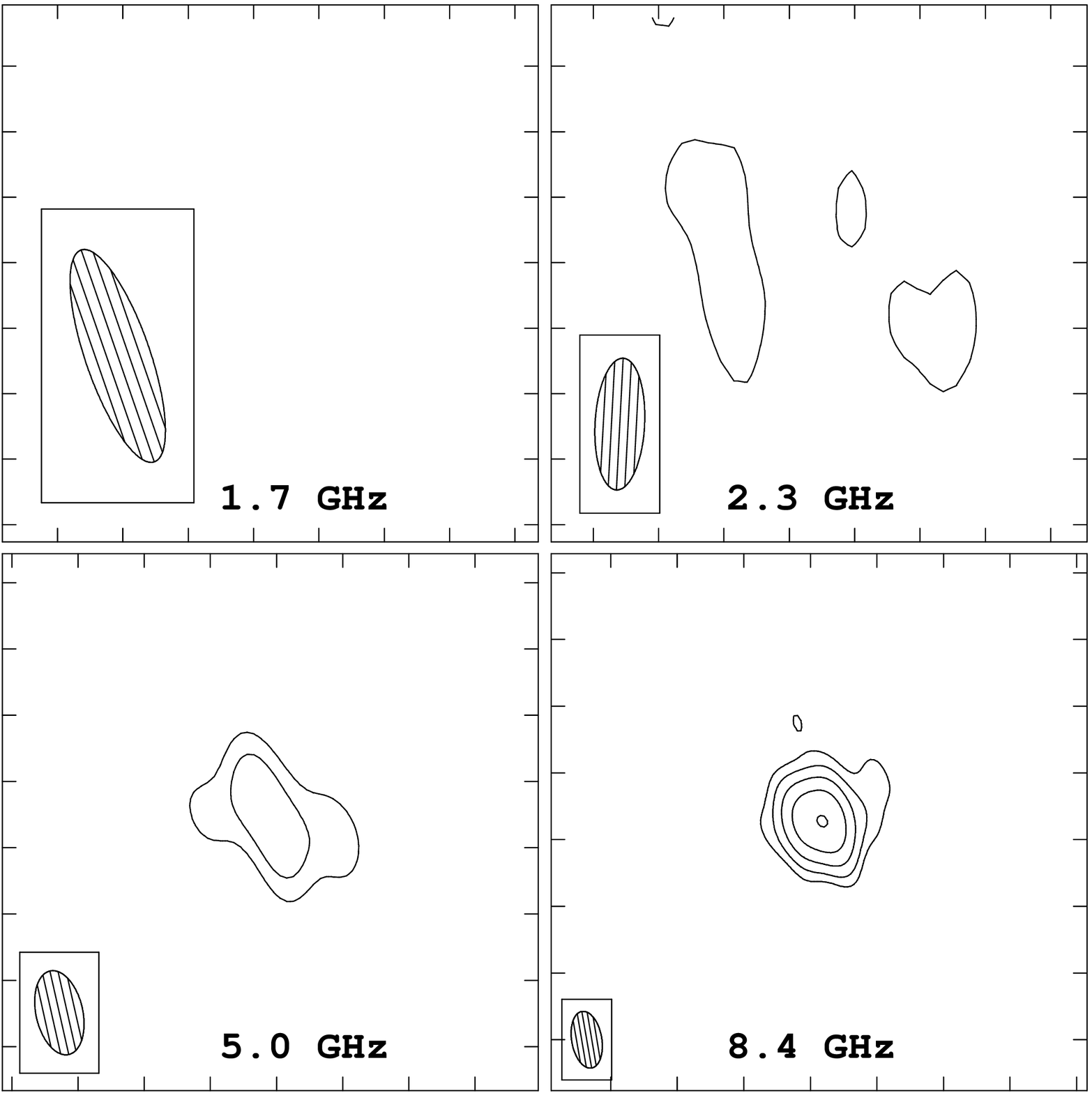}
\caption{Radio contour maps of the $0\farcs04\times 0\farcs04$ field
surrounding component E.  Contours begin at $3\sigma$ and increase by
factors of 2, where $\sigma$ is the R.M.S.\ noise level.  The
restoring beam is illustrated in the
lower left corner of each panel.
\label{fig:e}}
\end{figure}

\begin{figure}
\figurenum{7}
\epsscale{0.8}
\plotone{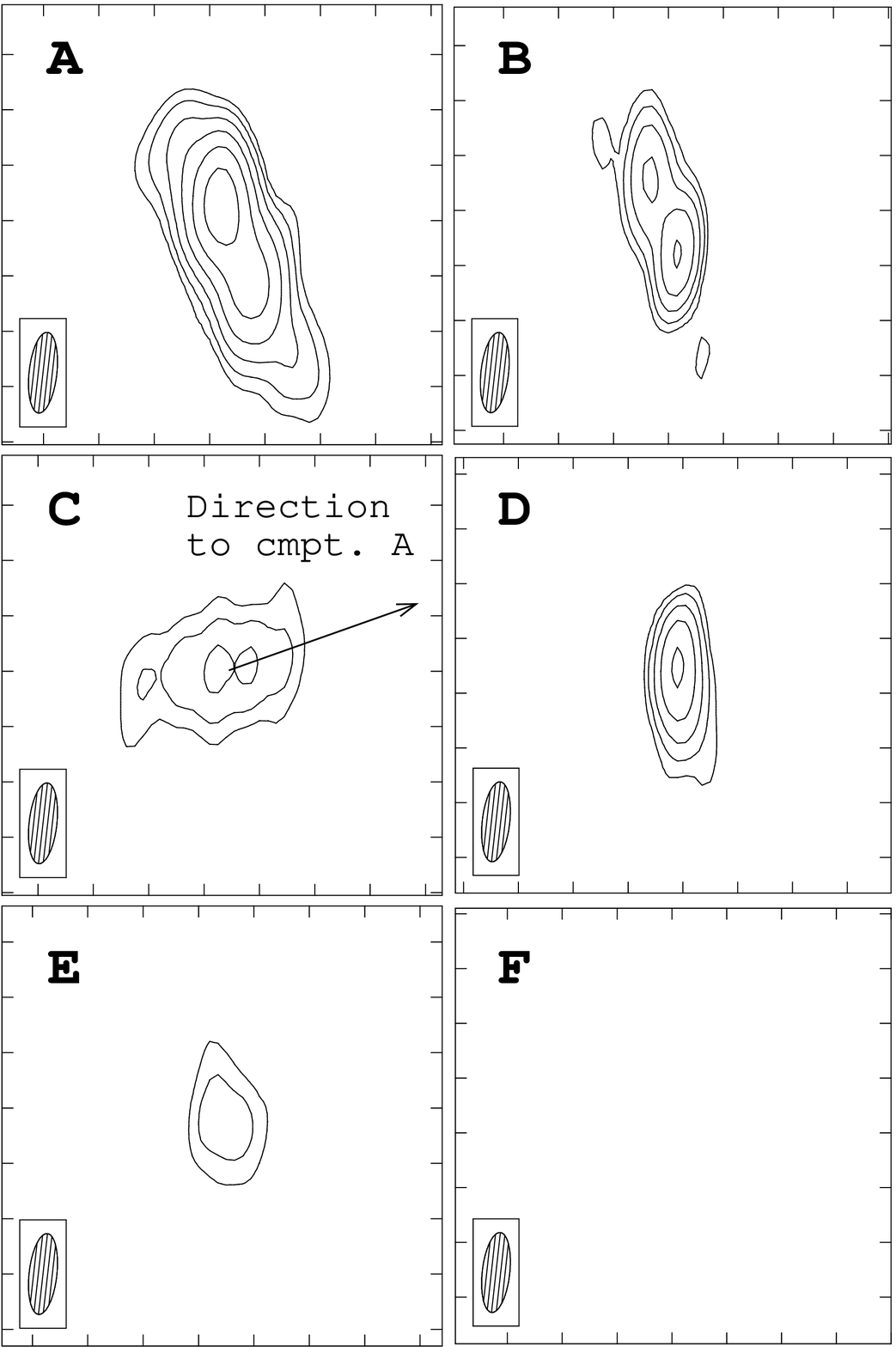}
\caption{15~GHz contour maps of the $0\farcs008\times 0\farcs008$
field surrounding each of components A--F.  Contours begin at
$3\sigma$ and increase by factors of 2, where
$\sigma=0.35$~mJy~beam$^{-1}$ is the R.M.S.\ noise level.  The
restoring beam is illustrated in the
lower left corner of each panel.
\label{fig:uband}}
\end{figure}

We also present narrow-field contour maps of the fields surrounding
components A, B, and C (Figure~\ref{fig:abc}), components D and F
(Figure~\ref{fig:df}), and component E (Figure~\ref{fig:e}).  The
15~GHz contour maps are presented separately for each component
(Figure~\ref{fig:uband}).  All the contour maps were created without
any tapering function and therefore employ the full resolution of the
observations.  These contour maps allow for a closer examination of
individual component sizes and shapes.  The most important parameters
describing these full-resolution images are given in
Table~\ref{tbl:journal}: the image dimensions, the major and minor
axes (full-width-half-maximum) and orientation of the restoring beam,
and the root-mean-squared (R.M.S.) noise level.

We draw attention to some features of these images and
contour maps that will be discussed in detail, and placed in the
context of the KW03 models, in subsequent sections: \begin{enumerate}
\item Component F is detected at all frequencies except 15~GHz.  It is
especially prominent at 1.7~GHz but fades relative to D as the
frequency increases (\S~\ref{sec:cmptf}).

\item Components C and E are heavily resolved at 1.7~GHz, are faint
and resolved at 2.3~GHz and 5~GHz, and are compact and bright at the
highest frequencies.  By contrast, components B and D are compact at
all frequencies (\S~\ref{sec:scatter}).

\item In the 1.7~GHz map, there is extended radio emission near the
expected location of C, but its centroid appears to be shifted
$\approx$30 mas south (\S~\ref{sec:counterimage}).

\item The radio arc between A and B was detected at the two lowest
frequencies.  In addition, at 1.7~GHz, there is a significant
extension of component A towards the expected location of C
(\S~\ref{sec:arc}).

\item At 15~GHz (Fig.~\ref{fig:uband}), component B appears to be a
fainter and parity-reversed copy of component A.  Component C is
elongated along a line pointing to A (\S~\ref{sec:arc}).

\end{enumerate}

\subsection{Component F}
\label{sec:cmptf}

\begin{figure}
\figurenum{8}
\epsscale{1.0}
\plotone{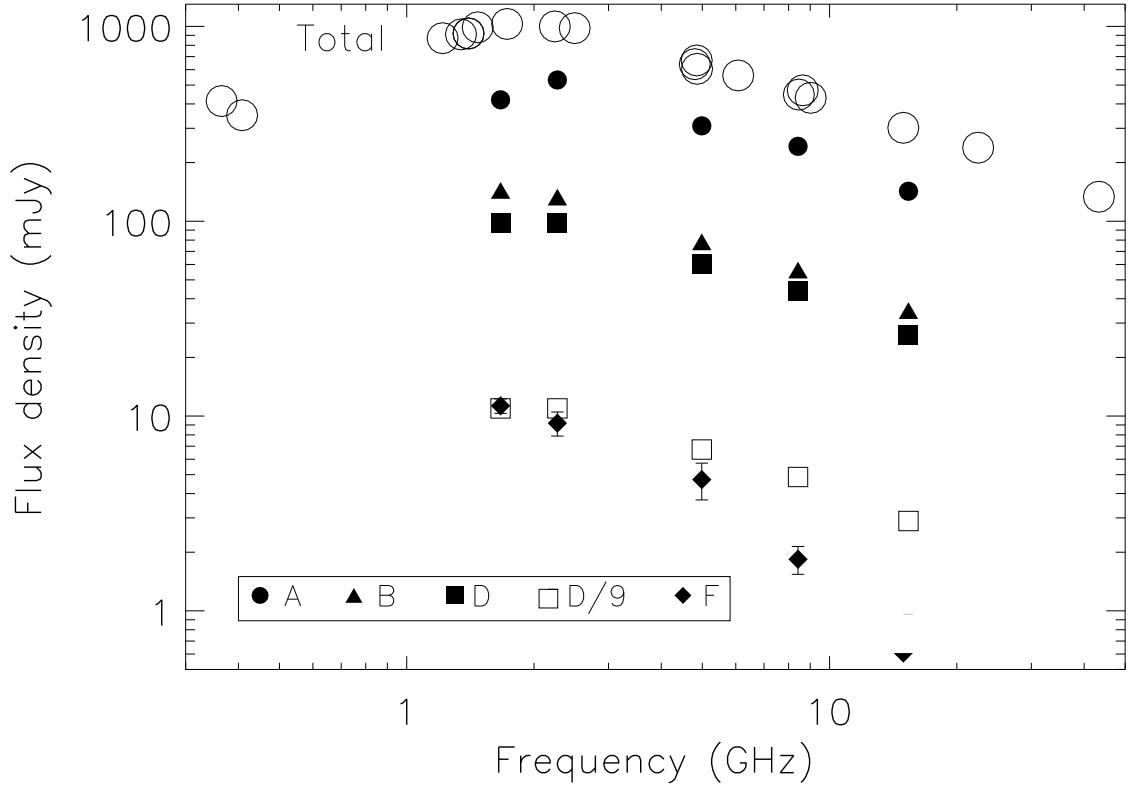}
\caption{Flux densities of components A, B, D, and F.  Where not shown
as error bars, the statistical errors are comparable to the symbol
sizes, or smaller.  Also plotted are the measurements of the total
flux density of all components, compiled by W02 from various sources.
\label{fig:cmptf}}
\end{figure}

We measured the total flux density of each radio component by fitting
an elliptical Gaussian function to the appropriate region of each VLBA
image using the {\sc aips} procedure {\sc jmfit}, which takes into
account the ellipticity of the restoring beam. The results are given
in Table~\ref{tbl:fluxes}.  The flux densities of components A, B, D,
and F are plotted in Figure~\ref{fig:cmptf} as a function of
frequency. Also plotted is the total flux density of all components,
from the measurements compiled by W02. The upper limit on the flux of
F at 15~GHz represents the $3\sigma$ level in that region of the
image. Results for components C and E are not plotted because, for
these components, a substantial amount of flux is ``resolved out'' at
low frequencies due to the interferometer's lack of short baselines.

The continuum spectra of components A, B, and D are nearly identical,
while the spectrum for component F is significantly steeper. To make
the difference in spectral slope obvious to the eye, we have divided
the flux densities of D by 9, to match the 1.7~GHz flux density of F,
and plotted the results with open squares.  Using the measurements at
which the components are fairly compact and the fitting results are
therefore most accurate, we find spectral slopes of $\alpha_{\rm A} =
-0.64\pm 0.08$, $\alpha_{\rm B} = -0.65\pm 0.07$, and $\alpha_{\rm D}
= -0.63\pm 0.07$ (where $S_\nu \propto \nu^{\alpha}$).  These values
agree with the value $\alpha=-0.69\pm 0.04$ measured by W02 using
lower-resolution data.  For component F, using measurements at all
frequencies for which it was detected, $\alpha_{\rm F} = -1.07\pm
0.13$.

In the KW03 scenario, A--E are multiple images of the same quasar, and
should have nearly identical spectral indices, but F need not have
(and probably would not have) the same spectral index.  It was already
established by W02 that A--E have the same spectral index.  By
establishing that F has a different spectral index than the other
components, our analysis has confirmed this key expectation of the
KW03 models.

\begin{deluxetable}{cccccc}
\tabletypesize{\scriptsize} \tablecaption{Total radio flux densities\label{tbl:fluxes}}
\tablewidth{0pt}
\tablehead{
\colhead{Component} & \multicolumn{5}{c}{Total flux density (mJy)} \\
\colhead{}          & \colhead{1.7~GHz} &  \colhead{2.3~GHz} &  \colhead{5.0~GHz} &  \colhead{8.4~GHz} &  \colhead{15~GHz}
}
\startdata
A & $\mathit{420\pm 20}$ & $\mathit{530\pm 30}$  & $309\pm 1$    & $242.4\pm 0.3$& $143\pm 1$\\
B & $\mathit{144\pm  7}$ & $\mathit{133\pm  7}$  & $78.4\pm 0.6$ & $56.2\pm 0.3$ & $35\pm 1$ \\ 
C & \nodata              & \nodata               & $18\pm 1$     & $41.4\pm 0.6$ & $28\pm 2$ \\ 
D & $98.5\pm 0.4$        & $98.5\pm 0.5$         & $60.6\pm 0.6$ & $43.8\pm 0.3$ & $26.0\pm 0.7$ \\ 
E & \nodata              & \nodata               & $8\pm 1$      & $12.3\pm 0.3$ & $12\pm 1$ \\ 
F & $11\pm 1$            & $9\pm 1$              & $5\pm 1$      & $1.8\pm 0.3$  & $< 0.96$ ($3\sigma$)
\enddata
\tablecomments{Flux densities estimated from elliptical Gaussian fits
to the radio components.  Italics indicate cases for which this
approximation is especially poor.  Entries for C and E are not given
for cases where the components were almost completely resolved out.}

\end{deluxetable}

\subsection{Scatter broadening}
\label{sec:scatter}

One objection to the KW03 models, or any scenario in which components
A--E are gravitationally lensed images of a single source component,
is that components C and E were observed to have a much lower surface
brightness at 1.7~GHz and 5.0~GHz than the other components. The
problem is that gravitational lensing cannot alter surface
brightness. However, it is possible for a non-gravitational effect,
namely scatter broadening due to electron-density fluctuations in the
lens galaxy or our own Galaxy, to cause differences in the surface
brightness of lensed images.  If different images pass through
different electron columns they will be broadened by different
amounts.

Scatter broadening can be recognized because the angular size of an
affected source increases strongly with observing wavelength. For a
source that is intrinsically a point source, the angular size due to
scatter broadening is typically proportional to $\lambda^n$ where
$n\approx2$ \citep[for a review, see][]{rickett90}. A particularly
well-studied case is Sgr~A$^{*}$ \citep[see,
e.g.,][]{jauncey89,lo93,doeleman01}. Among gravitational lenses, the
best-studied case of differential scatter broadening is PKS~1830--211
\citep{jones96,guirado99}. In that case, the broader image passes
through the spiral arm of the lens galaxy where a higher electron
column density might be expected \citep{winn02b,courbin02}.

Our VLBA maps strongly suggest that components A, C, and E are being
scatter broadened.  All three of these components are fairly compact
at 15~GHz, but are heavily resolved at 5~GHz and very heavily resolved
at 1.7~GHz.  This is despite the {\em decreasing} angular resolution
of the observations as one proceeds to lower frequencies.  To test the
$\lambda^2$-dependence, we fitted elliptical Gaussians to all the
components detected in each image (as in \S~\ref{sec:cmptf}).  The
fitting procedure takes into account the intrinsic ellipticity of the
beam and finds the deconvolved major and minor axes ($b_{\rm maj}$ and
$b_{\rm min}$).  We note that it would be preferable to fit models to
the visibility data rather than the image data, because the noise
properties of the visibility data are better understood, but in this
case visibility-fitting was unfeasible because of the large data
volume.  We also note that not all the components are well described
by ellipses, especially at 15~GHz, but the results from elliptical
fits should be accurate enough to determine the approximate
logarithmic scaling.

\begin{figure}
\figurenum{9}
\epsscale{1.0}
\plotone{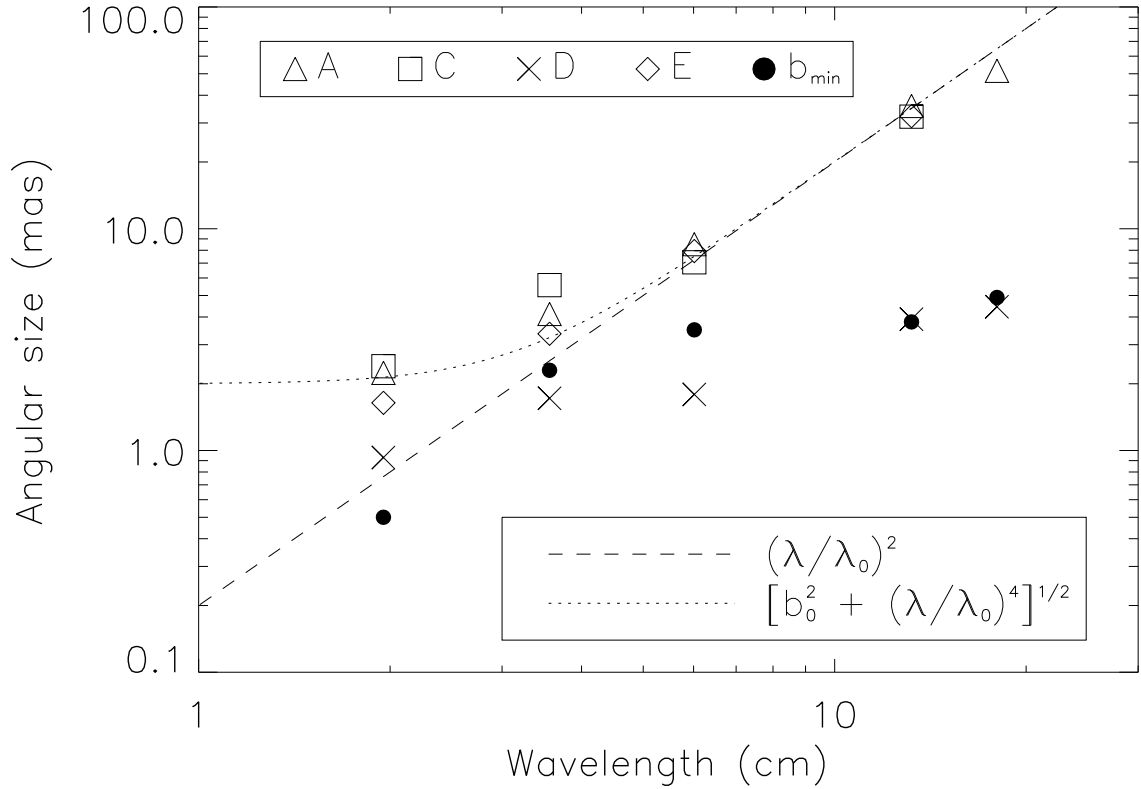}
\caption{Deconvolved major axes of components A, C, D, and E, as
measured from elliptical Gaussian fits to the VLBA images.  Black dots
indicate the size of the minor axis of the restoring beam.  The dashed
line is the trend expected for a scatter-broadened point source; the
dotted line is for a source with intrinsic size 2~mas.
\label{fig:scatter}}
\end{figure}

Figure~\ref{fig:scatter} shows the dependence of $b_{\rm maj}$ on
$\lambda$ for components A, C, and E.  As a measure of the angular
resolution, we have plotted black dots indicating the length of the
minor axis of the restoring beam, which is approximately proportional
to $\lambda$.  For comparison, we have also plotted the results for
component D, which does not exhibit scatter broadening; its size is
comparable to or smaller than the beam minor axis in all cases except
1.7~GHz.  (Component B does not appear to be scatter-broadened either,
but the results are not plotted in order to reduce clutter.)

Components A, C, and E are well-approximated by the $\lambda^2$ law
(dashed line), although the true dependence is not as steep,
especially when the 15~GHz (2~cm) point is included.  Given that
scatter broadening seems inevitable as an explanation for the
wavelength-dependent morphology of these components, we believe that
the departure from the $\lambda^2$ law shows that the observed angular
size is partially intrinsic at the shortest wavelengths.  The dotted
line shows the expectation of a scatter-broadened source with
intrinsic size 2~mas.  The source shapes in the 8.4~GHz and 15~GHz
images appear to be mainly intrinsic.  The flux densities, major and
minor axis sizes, and position angles of the components, taken from
the 15~GHz image in order to minimize scatter broadening, are given in
Table~\ref{tbl:shapes}.

The observed angular sizes can be used to estimate the scattering
measure (SM) along the lines of sight to the broadened components.
The SM is defined as the line-of-sight integral of $C_n^2$, where
$C_n^2$ is the normalization of the power spectrum of electron-density
fluctuations.  Assuming a Kolmogorov spectrum, the SM can be related
to the observed broadening of an extragalactic source via the relation
(see Fey, Spangler, \& Cordes 1991, Eq.\ 4; or Cordes \& Lazio 2003,
Eq.\ A17):
\begin{equation}
{\rm SM} = \left( \frac{\theta_{\rm FWHM}}{\rm 130~mas} \right)^{5/3}
           \left( \frac{\nu}{\rm 1~GHz} \right)^{11/3},
\end{equation}
where $\nu$ is the frequency at the lens redshift, and SM has the
units m$^{-20/3}$~kpc.  This predicts that angular size should vary as
$\lambda^{2.2}$, a little steeper than observed. Nevertheless, the
best fit to the data plotted in Fig.~\ref{fig:scatter} gives
SM~$\approx$~15--25~m$^{-20/3}$~kpc.  This value is too large to be
attributed to ionized material in our Galaxy.  The Galactic latitude
of \lens\ is $-70\arcdeg$, and the SM along such high-latitude lines
of sight in the Galaxy is typically less than
$10^{-3}$~m$^{-20/3}$~kpc (Cordes \& Lazio 2003, Fig.\ 6).

By contrast, the SM for low-latitude lines of sight in our Galaxy can
exceed $10^3$~m$^{-20/3}$~kpc, and has an average value of order unity
within $|b|<5\arcdeg$.  This suggests that a spiral lens galaxy viewed
nearly edge-on could produce the observed broadening, although it
seems unlikely that the disk would be aligned just right to intersect
all three components A, C, and E. It could also be that the lens
galaxy has a larger electron content than our Galaxy.  We caution,
however, that comparisons with our Galaxy may be misleading because
the observed major axes of $\approx$30~mas at 18~cm correspond to a
physical size of $\approx$150$h^{-1}$~pc at the lens redshift, which
is much larger than the length scales probed by scattering effects in
our Galaxy.  (For exactly that reason, observations of scatter
broadening in gravitational lens systems such as \lens\ may prove
useful for characterizing electron-density fluctuations on scales
inaccessible in our Galaxy.)

\begin{deluxetable}{cccccc}
\tabletypesize{\scriptsize} \tablecaption{Radio component shapes at
15~GHz\label{tbl:shapes}} \tablewidth{0pt} \tablehead{
\colhead{Component} & \colhead{Peak flux density} & \colhead{Total
flux density} & \colhead{Major axis} & \colhead{Minor axis} &
\colhead{Position angle} \\ \colhead{} & \colhead{(mJy~beam$^{-1}$)} &
\colhead{(mJy)} & \colhead{(mas)} & \colhead{(mas)} &
\colhead{($\arcdeg$~E of N)} }

\startdata
A &  $43.9\pm0.4$ & $143\pm2$ & $2.24\pm0.02$ & $0.48\pm0.03$ & $23.5\pm0.5$ \\
B &  $13.6\pm0.4$ & $ 35\pm1$ & $2.2\pm0.1$   & $0.1\pm0.1$   & $20\pm2$ \\
C &  $ 5.1\pm0.3$ & $ 28\pm2$ & $2.4\pm0.2$   & $0.8\pm0.3$   & $100\pm6$ \\
D &  $18.8\pm0.3$ & $ 26\pm1$ & $0.93\pm0.06$ & $0.2\pm0.1$   & $14\pm4$ \\
E &  $ 3.7\pm0.3$ & $ 12\pm1$ & $1.6\pm0.3$   & $1.0\pm0.2$   & $10\pm 20$
\enddata
\end{deluxetable}

\subsection{Counter-images of component F}
\label{sec:counterimage}

Component F, in the KW03 models, is actually the brightest of three
images of the second source component S$_2$.  The other two images of
S$_2$ (the ``counter-images'' of F) are expected to be located just
south of components C and E.  For convenience, we refer to the
hypothetical counter-images near C and E as components \cp\ and \ep,
respectively.  The KW03 models suggest that the flux ordering, from
brightest to faintest, is F--\cp--\ep. It is also possible that there
are two additional counter-images located between A and B, for a total
of five, but the models that produce these two additional
counter-images occupy a tiny volume in the phase space of model
parameters explored by KW03.

There is suggestive evidence for one of these counter-images in the
1.7~GHz map.  In Fig.~\ref{fig:abc}, black dots mark the
positions of A, B, and C (as measured at higher frequencies),
and the predicted position of \cp.  There is a
puff of extended radio emission located $\approx$30~mas due south of
the expected position of component C.  The location of the peak
emission is not consistent with C, but it is within 6~mas of the
predicted position of \cp.  Its total flux density is $\approx$10~mJy,
or $\approx$90\% of the total flux density of F, which is also the
expectation of the KW03 models.

It is therefore possible that C is almost completely resolved out at
1.7~GHz, and the extended emission that is observed is instead the
scatter-broadened counter-image \cp.  The reason we consider the
evidence ``suggestive'' rather than definitive is that \cp\ was not
detected at any other frequency.  In particular, the 8.4~GHz data are
inconsistent with $S_{{\rm C}'}/S_{\rm F}=0.9$, unless \cp\ is being
scatter-broadened even at that higher frequency.  Therefore, we
withhold final judgment on whether the detected flux is actually \cp\
until it can be detected at a higher frequency.  The challenge will be
to achieve the required signal-to-noise ratio in spite of the steep
spectrum of S$_2$ (see \S~\ref{sec:cmptf}).

Assuming that we did not detect additional radio components, we can
set upper limits on the fluxes on \cp\ and \ep.  We express these
limits as ratios relative to component F, because lens models directly
prescribe the magnification ratios of the counter-images relative to
component F.  We derive the limits from the 8.4~GHz map, because that
is the highest-frequency image in which F is detected.  By requiring
the flux density of a point source be below $5\sigma$, where $\sigma$
is the noise level of the 8.4~GHz image, we find $S_{{\rm C}'}/S_{\rm
F} < 0.57$ and $S_{{\rm E}'}/S_{\rm F} < 0.57$.

As discussed in detail by KW03, the limit on \ep\ is not constraining,
but the limit on \cp\ formally rules out all but a few models.
However, as mentioned previously, the true limit on \cp\ may be weaker
due to scatter broadening.  For example, assuming that the observed
sizes of C and E at 8.4~GHz are due entirely to scatter broadening, we
might allow \cp\ and \ep\ to have the same ratio of total to peak flux
density.  Modified in this way, the upper limits become $S_{{\rm
C}'}/S_{\rm F} < 2.1$ and $S_{{\rm E}'}/S_{\rm F} < 1.1$, which are
consistent with all the KW03 models.  The truth lies somewhere between
these extremes, because the observed sizes of C and E at 8.4~GHz are
probably at least partially intrinsic (see \S~\ref{sec:scatter}).

\subsection{Extended emission}
\label{sec:arc}

The radio arc joining components A and B was known previously, and is
a natural consequence of the KW03 models.  In addition, the KW03
models predict there should be a bridge of radio emission between
components A and C.  The detailed surface brightness profile of the
bridge depends on the source size and shape, but it is otherwise a
robust prediction.

There are two indications in the radio maps that this is correct.  The
first indication is found in the lowest-frequency (1.7~GHz) image, in
which there are faint traces of extended flux between components A and
C.  This can be seen in the upper left panel of
Fig.~\ref{fig:grayscale-maps}, as a faint prong extending eastward
from component A, but is more easily seen in the upper left panel of
Fig.~\ref{fig:abc}.  The lowest 3 contours ($3\sigma$, $6\sigma$, and
$12\sigma$) bulge eastward from A to C.

The second indication is the orientation of component C in the
higher-frequency (5~GHz, 8.4~GHz and 15~GHz) images.  The deconvolved
position angle of C is oriented nearly along the direction to
component A.  For example, at 15~GHz, where we expect the observed
orientation to be intrinsic, component C is elongated with position
angle $100\arcdeg\pm 6\arcdeg$, whereas the direction from C to A has
position angle $110\arcdeg$ (see the lower left panel of
Figure~\ref{fig:abc}).

\subsection{Component shapes}
\label{sec:shapes}

Since we have argued that the observed sizes of the radio components
at 15~GHz are mainly intrinsic, the morphologies of the components
should provide clues about the properties of the background source and
the correct lens model.  In Fig.~\ref{fig:uband}, component A appears
to be a two-sided core-jet source. There is a longer and more
prominent jet directed southwest of the core, towards B, and there is
a shorter jet directed northeast.  Component B also appears to be a
core-jet source, but only one jet is observed, and is directed
northeast, towards A.  Apparently, A and B are opposite-parity images
of a two-sided core-jet source, but only one jet is apparent in B due
to the relative compression of that image by the lens mapping.

The morphologies of the other components are simpler.  Component C is
highly elongated towards A, as already noted.  Component D is mainly
compact but does have a faint extension to the southwest, suggesting
that D and A have the same parity.  Component E is resolved but is too
faint for a detailed examination.  Component F was not detected.

The KW03 models predict that A and B have opposite parities, and that
A and D have the same parity, which is consistent with the observed
morphologies.  In addition, the position angles given in
Table~\ref{tbl:shapes} for components A--E can be produced by 23\% of
the models discussed by KW03.  However, we caution against
over-interpreting the shape parameters.  As described in
\S~\ref{sec:scatter}, they were derived by the simple procedure of
fitting a single elliptical Gaussian function to each component.  The
quoted errors are the statistical errors in the fitting process, and
are therefore under-estimated for those components that are not well
described by ellipses (especially A and B). The most secure parameters
are the position angles, and the least secure parameters are the minor
axis lengths (which in some cases are considerably smaller than the
restoring beam).

\section{Optical evidence}

\subsection{Observations and data reduction}
\label{sec:optical}

\begin{figure}
\figurenum{10}
\epsscale{1.0}
\plotone{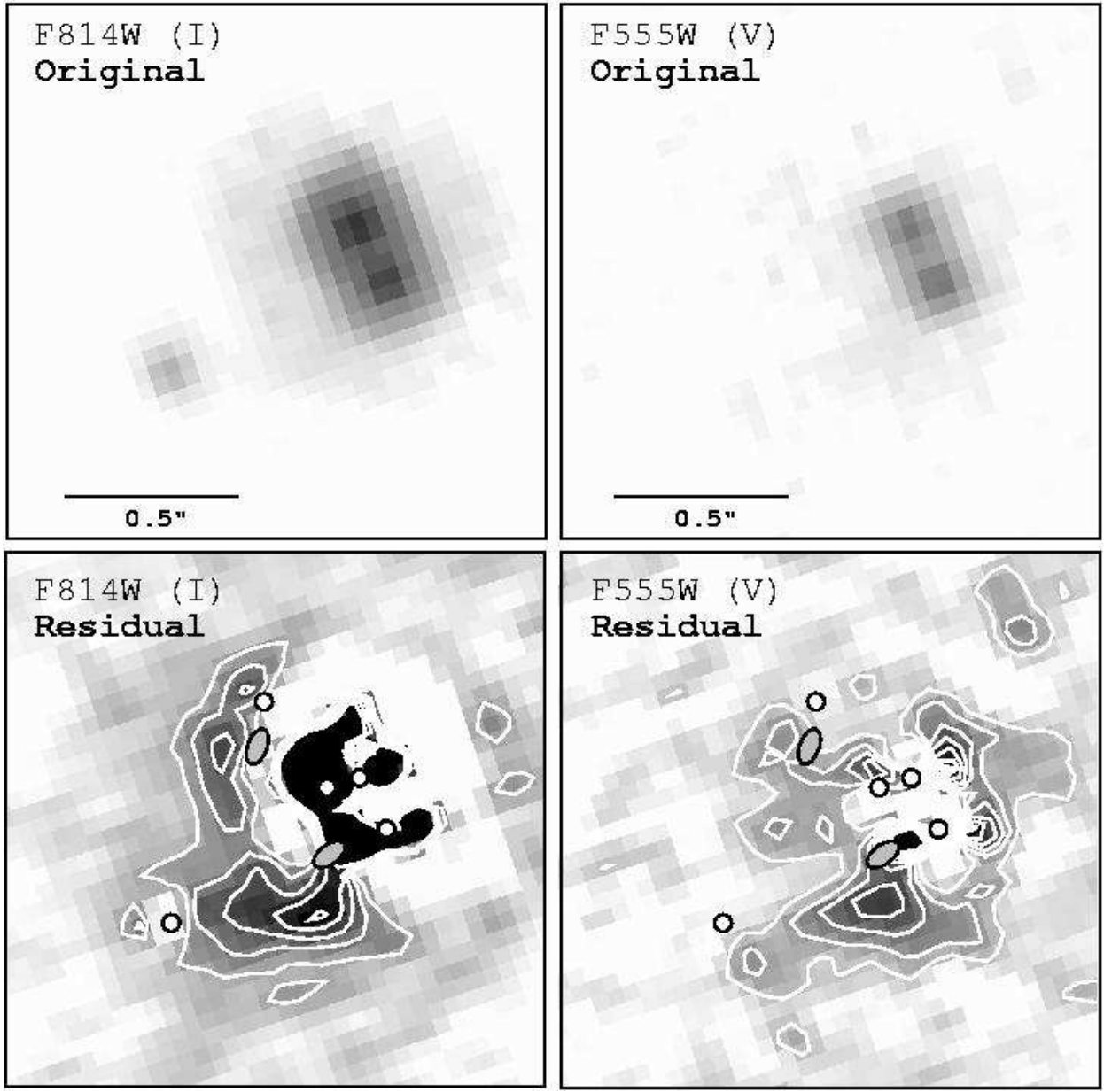}
\caption{HST images of \lens.  North is up and east is left.  In the
residual images (bottom), the positions of the subtracted components
A--E are marked by white circles. The small gray ellipses indicate the
approximate range of lens galaxy positions predicted by KW03.
\label{fig:hst}}
\end{figure}

On 2001~July~18 we obtained optical images of \lens\ in the F555W
(hereafter, $V$) and F814W ($I$) filters on the WFPC2 camera of the
HST.  We obtained two 21.7-minute exposures at each of two dither
positions, for a total exposure time of 86.8~minutes per filter.  The
target was centered in the PC chip, which has a pixel scale of
$0\farcs0456$~pixel$^{-1}$.  The exposures were combined and cosmic
rays rejected using standard {\sc iraf}\footnote{The Image Reduction
and Analysis Facility ({\sc iraf}) is a software package developed and
distributed by the National Optical Astronomical Observatory, which is
operated by {\sc aura}, under cooperative agreement with the National
Science Foundation.} procedures.  The images are shown in the upper
two panels of Figure~\ref{fig:hst}.  Components A and B were detected
in both filters.  Component D was detected in $I$-band but is nearly
absent in $V$-band.

These same data were previously analyzed by \citet{hall02}, who also
measured the optical flux ratios and detected residual flux after
subtracting point sources at the positions of the radio components.
Our results confirm this previous work.  However, as described below,
our more detailed methods have allowed us to proceed further in the
interpretation of the results.

\subsection{Detection of the lens galaxies}
\label{sec:lensgals}

We built photometric models of the HST images using custom software
for convolving a parametric model with a simulated PSF and varying the
model parameters to achieve the best fit to the images. The simulated
PSF was generated with the Tiny Tim program by \citet{tinytim}. For
the $V$-band image, the model consisted of 5 point sources,
representing the radio components A--E.  At $I$-band we found that
components A and B are not well represented by point sources, possibly
due to a contribution from the quasar host galaxy.  We obtained much
smaller residuals by allowing for small Gaussian components at the
locations of both A and B, in addition to point sources.

Even with this modification, there were still significant residuals.
In Figure~\ref{fig:hst} we show the residuals as grayscale images
overlaid with white contours.  The positions of the subtracted
components are marked with white circles.  The ring of negative
residuals surrounding component A is spurious and due to imperfect PSF
subtraction, but there are two peaks of positive emission in the
residual images.
\begin{enumerate}
\item The ``northern peak,'' $0\farcs14$ south and $0\farcs12$ east of
E, and
\item The ``southern peak,'' $0\farcs37$ south and $0\farcs03$ east of
C.
\end{enumerate}

We believe these two peaks represent genuine emission rather than
artifacts of the subtraction.  This pattern of residuals did not
appear when we applied the same fitting and subtraction procedure to
an isolated field star on the PC chip.  In addition, although the
peaks are more prominent in $I$-band, they are detected in both
filters, despite the completely different PSF.

\citet{hall02} already noted the presence of low-surface-brightness
flux underlying the point sources.  However, the pattern in their
residual image is less obvious, possibly due to an inferior PSF
subtraction.  They used only integral pixel shifts, did not
simultaneously fit for the fluxes of the point sources, and did not
allow for any extension of A and B.

Interestingly, G02 also found two residual peaks of emission by
deconvolving a ground-based $K'$-band image.  The locations of each
$K'$-band peak is within $0\farcs2$ of the corresponding peak in our
images, which is fairly good agreement, considering that the spurious
residuals around A may be corrupting our measurements of the peak
locations.  Although G02 were unsure whether the $K'$-band peaks
represented genuine emission, or artifacts of the deconvolution, we
argue that the observation of similar peaks at both $I$-band and
$V$-band lends confidence to all three detections.

The detection of two diffuse light sources is a further confirmation
of the two-galaxy KW03 lens models.  As discussed further by KW03, the
predicted positions of the galaxies are both within $0\farcs15$ of the
observed locations at $I$-band and $V$-band. The approximate range of
lens galaxy positions predicted by KW03 is indicated by the two small
gray ellipses in Figure~\ref{fig:hst}.

It is not clear how seriously to take the observed morphologies of the
northern and southern peaks, because they are faint, and because of
possible contamination by spurious residuals from the PSF subtraction.
Nevertheless, given the morphology of the southern peak observed in
both filters, we conjecture that the southern lens galaxy is elongated
east-west, and overlaps component D at its easternmost extent.  This
would explain the very red color of D (see \S~\ref{sec:colors}).

\subsection{Photometry and colors of the components}
\label{sec:colors}

We measured the optical fluxes of components A--E using the
PSF-fitting procedure described above.  The optical flux ratios for
A--E, relative to component B, are given in Table~\ref{tbl:ratios} for
both the $I$ and $V$ filters.  For some components we could derive
only upper limits.  For reference, we also list the radio values
measured by W02.

\begin{deluxetable}{cccc}
\tabletypesize{\scriptsize}
\tablecaption{Flux ratios relative to component B\label{tbl:ratios}}
\tablewidth{0pt}
\tablehead{
\colhead{Component} &
\colhead{F555W$\approx V$} &
\colhead{F814W$\approx I$} &
\colhead{Radio}
}

\startdata
A & $0.86\pm 0.05$ & $1.65\pm 0.09$   &  $4.1\pm 0.2$ \\
B & $1.00$         & $1.00$           &  $1.00$ \\
C & $<0.18$        & $<0.18$          &  $0.87\pm 0.05$ \\
D & $<0.011$       & $0.045\pm 0.003$ &  $0.76\pm 0.03$ \\
E & $<0.017$       & $<0.004$         &  $0.24\pm 0.01$
\enddata
\end{deluxetable}

To convert the fluxes to standard photometric magnitudes, we adopted
zero points of 21.69 for F814W and 22.56 for F555W, based on the work
of \citet{dolphin00} and corrected for infinite aperture.  No
correction was attempted for the charge-transfer-efficiency (CTE)
problem of WFPC2.  The total magnitudes of all the components in our
model was $I=18.84\pm 0.05$ and $V=22.54\pm 0.07$, in good agreement
with the ground-based photometry of W02 ($I=18.79\pm 0.05$,
$V=22.47\pm 0.09$).

All the components that were detected are very red.  Even component B,
the bluest of the detected components, has $V-I=3.4$.  This is redder,
for example, than any of the 157 quasars in the radio-selected (and
optically unbiased) sample of Parkes sources that was investigated by
\citet{fww00} (see Fig.\ 4 of that work).

The optical flux ratios disagree strongly with the radio flux ratios,
even though gravitational lensing is achromatic.  This is a common
observation among gravitational lenses, usually attributed to dust
extinction by the lens galaxy \citep[see, e.g.,][]{falco99} or optical
microlensing \citep{wp91}, although in this case the differential
extinction is larger than usual.  By comparing the radio flux ratios
and the optical flux ratios, we can estimate the difference between
the $I$-band extinction $A_I$ incurred by B (the bluest component),
and the $I$-band extinction incurred by the other components:
\begin{equation}
   \Delta A_I({\rm A}) =
   A_I({\rm A}) - A_I({\rm B}) = -2.5\log_{10}
   \left(
         \frac{f_I({\rm A})/f_I({\rm B})}
              {f_{\rm radio}({\rm A})/f_{\rm radio}({\rm B})}  \right),
\end{equation}
where $f_I({\rm B})$ refers to the $I$-band flux of component B, and
so forth.  The results for the components, in order from bluest to
reddest, are: $\Delta A_I({\rm B}) \equiv 0$, $\Delta A_I({\rm
A})=0.97$, $\Delta A_I({\rm C}) >1.7$, $\Delta A_I({\rm D}) =3.1$, and
$\Delta A_I({\rm E}) >4.4$. (The position of C in the ordering is
uncertain because the result is a relatively weak upper limit.)

The most plausible explanation for the different optical colors of all
the lensed components is differential reddening due to dust in the
lens galaxies. The main alternative explanation, microlensing, is
implausible because it is unlikely that microlensing would affect more
than one image to the required degree at the same time.  \citet{wt02},
for example, estimate there is only a 30\% chance for even one image
of a multiple-image quasar to vary by more than 0.5~mag over
10~years. In addition, the dust hypothesis explains why the fainter
components are also redder; this would be a coincidence under the
microlensing hypothesis.  Therefore, at least part of the reason for
the extreme red color of \lens\ is dust extinction in the lens
galaxies. This need not be the entire reason. Because we can determine
only the differential extinction, rather than the total extinction, we
cannot rule out the existence of dust in the host galaxy that reddens
the background source, or an intrinsically red background source.

\section{Summary and conclusions}
\label{sec:summary}

The evidence presented in this paper strongly supports the models
developed in detail by \citet{kw03}: the unique radio morphology of
\lens\ represents five images (A--E) of a single radio-loud
quasar. There is also a sixth radio component (F) representing a
different part of the background radio source.

Radio observations with the VLBA at five widely spaced frequencies
demonstrate that F has a steeper spectral index than components A--E,
ruling out six-image models and supporting the notion that F
represents a different part of the background radio source.  One of
the predicted counter-images of F may have been detected at 1.7~GHz,
but this needs to be confirmed with more sensitive observations at
higher frequencies.  If there remains any doubt about the KW03
scenario, detection of both of the counter-images would be definitive.

The multi-frequency maps also demonstrate that the lower surface
brightness of components A, C, and E at long wavelengths is due to
differential scatter broadening.  This defends the theory that they
are multiple images of a single quasar from the objection that surface
brightness must be conserved by gravitational lensing.  Components B,
D, and F do not appear to be significantly scatter broadened.

At short wavelengths, where scatter broadening is minimized, the
resolved shapes of the radio components can be broadly explained by a
subset of the KW03 models.  In particular, component C is aligned in
the direction of component A, as predicted generically by those
models.  The radio arc between components A and B, which was
previously detected at 1.7~GHz and which we also detected at 2.3~GHz,
emerges naturally from the models.  At the longest wavelength there is
a trace of extended emission between A and C, again fulfilling the
expectation of the lens models that these components should be
connected by an arc.

The positions of the two lens galaxies proposed by KW03 are in rough
agreement with two residual peaks in optical images with the HST, and
also with two residual peaks identified by G02 in a deconvolved
$K'$-band image.  The northern galaxy appears to be responsible for
the reddening and scatter broadening observed in components A, C, and
E.  The southern galaxy appears to be responsible for the reddening of
component D.  Component B, which is the most distant in projection
from either galaxy, is not scatter broadened and is the bluest
component, as expected.  It should be possible to improve the
characterization of the lens galaxies with space-based images, or
ground-based images with a very high signal-to-noise ratio (allowing a
reliable deconvolution).  Given the presence of dust and ionized
material in the lens galaxies and the possibly elongated morphology
observed for the southern lens galaxy, it seems likely that both
galaxies are spiral galaxies.

\acknowledgments We thank Ramesh Narayan and David Rusin for useful
discussions.  We are indebted to Steven Beckwith for providing
Director's Discretionary Time on the HST for this project, and we
thank Emilio Falco, Chris Impey, Brian McLeod, and Hans-Walter Rix,
for helping to obtain the HST data as members of the {\sc castles}
collaboration.  J.N.W.\ is supported by an NSF Astronomy \&
Astrophysics Postdoctoral Fellowship, under grant AST-0104347. C.S.K.\
is supported by {\sc nasa} ATP grant NAG5-9265 and HST grant GO-9133.


\begin{thebibliography}{}

\bibitem[Cordes \& Lazio(2003)]{cl03} Cordes, J.M.\ \& Lazio, T.J.W.\
2003, preprint [astro-ph/0301598]

\bibitem[Courbin et al.(2002)]{courbin02} Courbin, F., Meylan, G.,
Kneib, J.-P., \& Lidman, C.\ 2002, \apj, 575, 95

\bibitem[Doeleman et al.(2001)]{doeleman01} Doeleman, S., et al.\
2001, \aj, 121, 2610

\bibitem[Dolphin(2000)]{dolphin00} Dolphin, A.E.\ 2000, \pasp, 112,
1397

\bibitem[Fey, Spangler, \& Cordes(1991)]{fsc91} Fey, A.L., Spangler,
S.R., \& Cordes, J.M.\ 1991, \apj, 372, 132

\bibitem[Falco et al.(1999)]{falco99} Falco, E., et al.\ 1999, \apj,
523, 617

\bibitem[Francis, Whiting, \& Webster(2000)]{fww00} Francis, P.J.,
Whiting, M.T., \& Webster, R.L.\ 2000, Publ.\ Astron.\ Soc.\ Aust.,
53, 56

\bibitem[Gregg et al.(2002)]{gregg02} Gregg, M., et al.\ 2002, \apj,
564, 133 (G02)

\bibitem[Guirado et al.(1999)]{guirado99} Guirado, J.C., et al.\ 1999,
\aap, 346, 392

\bibitem[Hall et al.(2002)]{hall02} Hall, P., et al.\ 2002, \apj, 575,
L51

\bibitem[Jauncey et al.(1989)]{jauncey89} Jauncey, D.L., et al.\ 1989,
\aj, 98, 44

\bibitem[Jones et al.(1996)]{jones96} Jones, D., et al.\ 1996, \apj,
470, L23

\bibitem[Keeton \& Winn(2003)]{kw03} Keeton, C.R.\ \& Winn, J.N.\
2003, preprint (KW03)

\bibitem[Krist \& Hook(1997)]{tinytim} Krist, J.E.\ \& Hook, R.N.\
1997, The TinyTim User's Guide, version 4.4 (Baltimore: STScI)

\bibitem[Lo et al.(1993)]{lo93} Lo, K.Y., et al.\ 1993, Nature, 362,
38

\bibitem[Rickett(1990)]{rickett90} Rickett, B.J.\ 1990, \araa, 28, 561

\bibitem[Rusin et al.(2001)]{rusin01} Rusin, D., et al.\ 2001, \apj,
557, 594

\bibitem[Wambsganss \& Paczynski(1991)]{wp91} Wambsganss, J.\ \&
Paczynski, B.\ 1991, \aj, 102, 864

\bibitem[Winn et al.(2002a)]{winn02} Winn, J.N., et al.\ 2002a, \apj,
564, 143 (W02)

\bibitem[Winn et al.(2002b)]{winn02b} Winn, J.N., et al.\ 2002b, \apj,
575, 103

\bibitem[Wyithe \& Turner(2002)]{wt02} Wyithe, J.S.B.\ \& Turner,
E.L.\ 2002, \apj, 575, 650

\end{thebibliography}
\end{document}